\documentclass[12pt,preprint]{aastex}
\notetoeditor{refer to: Fiamma Capitanio. Institute: IASF-INAF,Via Fosso de
l Cavaliere, 100, 00133 Rome ITALY. Tel: +390649934557. E-mail: fiamma.capitanio@rm.iasf.cnr.it ; pietro.ubertini@rm.iasf.cnr.it}
\newcommand{\inst}{\altaffilmark}

\shorttitle{Spectral states of the new X-ray binary IGR J17091--3624}
\shortauthors{Capitanio F. et al.}
\begin{document}
   \title{Spectral states of the X-ray binary IGR J17091--3624 observed by INTEGRAL and RXTE}

   \author {F. Capitanio\inst{1,}\inst{3}, A. Bazzano\inst{1}, P. Ubertini\inst{1}, A. A. Zdziarski\inst{2}, A. J. Bird\inst{3}, G. De Cesare\inst{1}, A. J. Dean\inst{3}, J.  B. Stephen\inst{4}, A. Tarana\inst{1} }

\altaffiltext{1}{IASF/INAF, Rome, Italy, Via Fosso del Cavaliere 100, I-00133 Rome, Italy}
 \altaffiltext{2}{Centrum Astronomiczne im.\ M. Kopernika, Bartycka 18, 00-716 Warszawa, Poland}
 \altaffiltext{3}{School of Physics and Astronomy, University of Southampton, Highfield Southampton, SO17 1BJ, UK}
 \altaffiltext{4}{ IASF/INAF, Bologna, Via Godetti 101, I-40129 Bologna, Italy}
              

\begin{abstract} IGR J17091--3624 was discovered  in 2003 April by {\it 
INTEGRAL}/IBIS during its Galactic Centre Deep Exposure programme. The source was initially detectable only in the 40--100 keV range, but after two days was also detected in the 15--40 keV range. Its flux had by then increased to 40 mCrab and 25 mCrab in the 15--40 keV and 40--100 keV  bands respectively. {\it RXTE\/} observed the source simultaneously on 2003 April 20, with an effective exposure of 2 ksec. We report here the spectral and temporal evolution of the source, which shows a transition between the hard and soft states. We analyse in detail the {\it RXTE}/{\it INTEGRAL\/} Comptonised spectrum of the hard state as well as the JEM-X detection of a blackbody component during the source softening. Even though the source spectral behavior and time variability show a similarity with the outburst of the black-hole candidate IGR J17464--3213 (= H1743--322), observed by {\it INTEGRAL\/} in 2003, the nature of its compact object (BH vs.\ NS) remains controversial. 
\end{abstract}


\section{Introduction}

A new source was discovered by {\it INTEGRAL}/IBIS during a pointed observation on 2003  April 14--15. It was designated IGR J17091--3624 based on its position of R.A.\ (2000) $=17^{\rm h}09.1^{\rm m}$, Dec.\ $=-36\degr 24.5\arcmin$, with an error box of $3\arcmin$ at 90\% confidence \cite{Atel1}. 

Initially, its flux was $\sim$20 mCrab in the 40--100 keV energy band exhibiting 
a hard spectrum, while it was not detected in the 15--40 keV band with an upper 
limit of $\sim$10 mCrab. A week earlier, the source did not appear in either 
energy band, with upper limits of $\simeq$10 mCrab \cite{Atel1}. During 
subsequent observations of the Galactic Center Deep Exposure (GCDE) on 2003 
April 15--16, the source flux  increased to $\sim$25 mCrab in the 40--100 keV 
band. The source was also included in the {\it INTEGRAL\/} IBIS/ISGRI Galactic 
plane survey catalogue \cite{Bird} and classified as unknown. 
Observations with the Very Large Array revealed the presence of a  
possible radio counterpart with a steeply falling spectrum typical of 
synchrotron emitters \cite{Atel3}, this was also confirmed by a subsequent observation 
with the Giant Metrewave Radio Telescope (Pandey et al.\ 2006).

Immediately after the {\it INTEGRAL\/} discovery, an {\it RXTE\/} observation 
was performed. The analysis of the {\it RXTE\/} data showed a relatively stable 
source flux without features associated with quasi-periodic oscillations (QPOs) 
in the power spectrum. Based on the energy and power spectra, Lutovinov \& 
Revnivtsev (2003) suggested that IGR J17091--3624 is an X-ray binary in the 
low/hard state and probably a black-hole system. IGR J17091--3624 was then 
searched for in the X-ray   catalogs and detected in the archival data of both 
the TTM telescope on board the KVANT module of the {\it Mir\/} orbital station 
\cite{Atel2}, and in the {\it BeppoSAX\/} WFC \cite{Atel4}. Following this, 
Lutovinov et al.\ (2005) performed a preliminary study of the IBIS/ISGRI 
spectrum of the source, showing that its flux changed by a factor of $\sim$2 
between 2003 April and 2003 August--September. During 2003, the source spectrum 
was described by a simple power law with a photon index of $\Gamma\simeq 2.2$ in 
the 20--150 keV energy band, i.e., softer than the {\it RXTE\/} spectrum. As the 
source flux in the hard X-rays increased in 2004, the spectrum became harder 
with $\Gamma\simeq 1.6$. From the above investigations, IGR J17091--3624 appears 
as a moderately bright variable (probably a transient) source with flaring 
activity in 1994 October ({\it Mir}/KVANT/TTM), 1996 September ({\it 
BeppoSAX}/WFC), 2001 September ({\it BeppoSAX}/WFC, in 't Zand et al.\ 2003), 
and 2003 April ({\it INTEGRAL}/IBIS, Kuulkers et al.\ 2003).

\subsection{Improvements over previous analysis}

We have performed a more accurate spectral analysis of this source compared to those
previously reported, analysing all the {\it INTEGRAL\/} data available 
while the source was detectable (until the end of 2004 April). The new set of 
data allowed us to fit a joint {\it RXTE}/{\it INTEGRAL\/} spectrum of IGR 
J17091--3624 in the low/hard state. The improved version of the {\it INTEGRAL\/} 
data analysis software (OSA 4.2, Goldwurm et al.\ 2003) has allowed us to obtain 
{\it INTEGRAL\/} spectra up to 150 keV. An initial test with the newly delivered OSA 5 
software has shown a substantial agreement with the former OSA 4.2 results. 

We have fitted the data with detailed models, thus obtaining new information on 
the geometry of the source and on the accretion disk optical depth and 
temperature. We have also extracted a JEM-X spectrum that confirms the previous 
hypothesis of a transition to the soft state \cite{Lut} and furthermore we have 
found suggestions of a hysteresis-like behaviour in the source spectral 
evolution, with the hard-to-soft state transition occurring at higher luminosity 
than the soft-to-hard transition. We then speculate on the source nature, even 
if the final picture is still not evident.   

\section{Data Analysis}

In this paper, we analyse the data set consisting of all {\it INTEGRAL\/} 
\cite{Winkl} Core Program and public observations in which IGR J17091--3624 was 
within the IBIS field of view, and of the {\it RXTE\/} Proportional Counter 
Array (PCA, Bradt et al.\ 1993) data collected during the 2 ks observation 
performed on 2003 April 20. The {\it INTEGRAL\/} observations, though not 
continuous, cover a period of one year from 2003 April to 2004 April. The {\it 
RXTE\/} observation corresponds to a period between {\it INTEGRAL\/} rev.\ 61 
(2003 April 14--16) and the beginning of rev.\ 63 (2003 April 21). In general, 
all {\it INTEGRAL\/} observations are divided into uninterrupted 2000-s time 
intervals, called science windows (SCW). Table~\ref{integral_log} summarises the 
{\it INTEGRAL\/} observations of the source. Light curves, hardness ratio and 
spectra are then obtained for each individual SCW. We use here {\it INTEGRAL\/} 
data from the coded mask imager, IBIS \cite{Uber}. Then, in order to obtain 
broad-band spectra, we add the lower energy data from the PCA, and, whenever 
available, those from the {\it INTEGRAL\/} X-ray Monitor, JEM-X \cite{Lund}. 

\begin{figure}
   \centering
 \includegraphics[angle=0, scale=0.4]{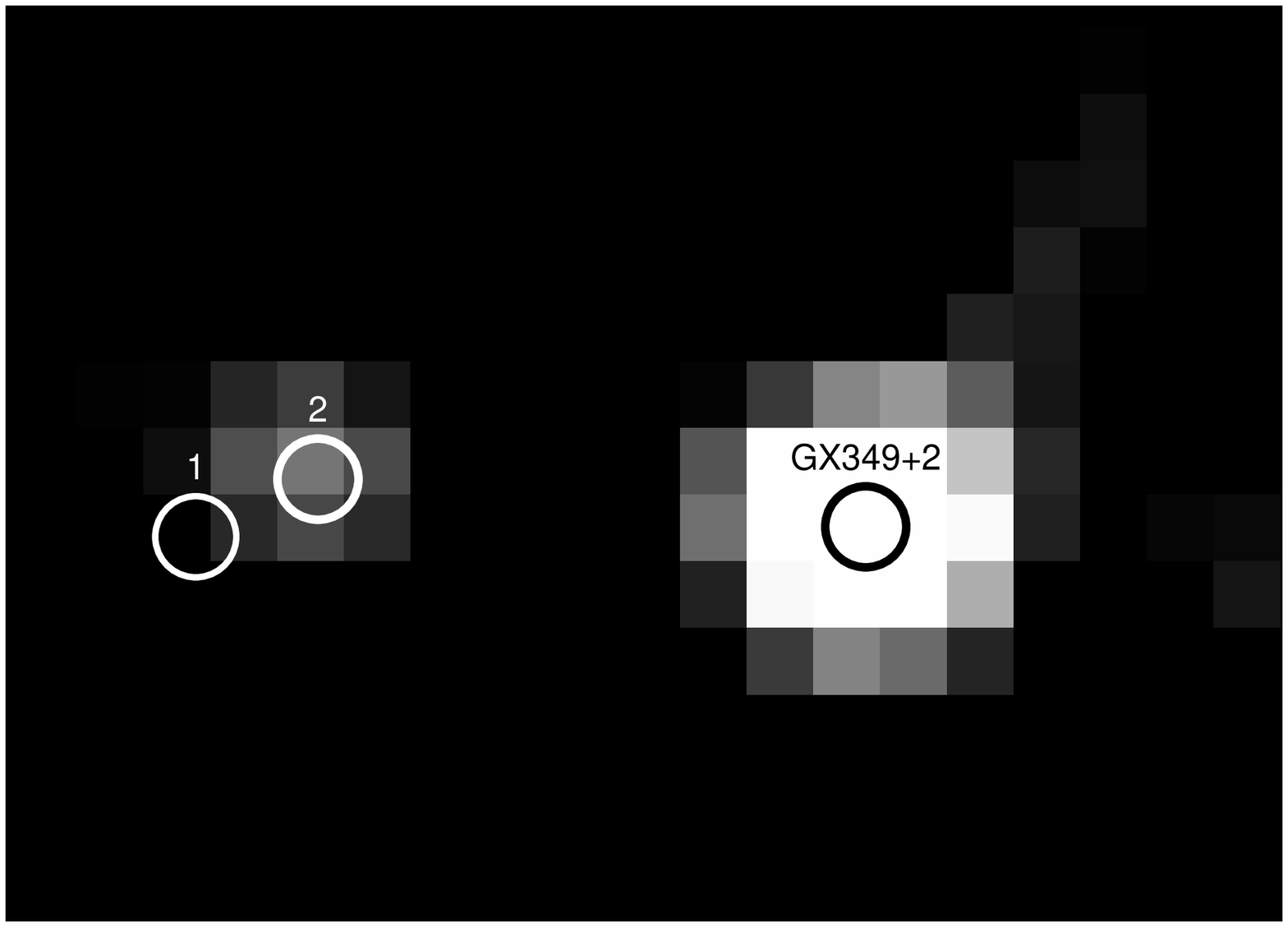}
      \caption{The 20--150 keV IBIS image of the field near IGR J17091--3624. Can be seen that IGR J17098-3628 (1) is not visible during the {\it RXTE}/PCA observations of IGR J17091--3624 (2), which argues against any contamination in the PCA field of view.}
         \label{image}
\end{figure}

Figure \ref{image} shows the 20-150 keV {\it INTEGRAL\/} image of IGR 
J17091--3624 during the period of the {\it RXTE\/} observation. The figure shows 
that the transient source IGR J1709.8-3628, detected $9.4\arcmin$ from IGR 
J17091--3624 on 2005 March 24 2005 \cite{source}, was then not visible. This 
rules out any contamination of other sources in the {\it RXTE}/PCA field of 
view.

The IBIS data were processed using the Off-line Scientific Analysis, OSA 4.2, 
Goldwurm et al.\ 2003) released by the {\it INTEGRAL\/} Science Data Centre, 
ISDC \cite{Courv}. The data set has been divided in two subsets according to the 
source position being either within the partially coded field of view (PCFOV, 
$19\degr\times 19\degr$) or in the narrower fully coded field of view (FCFOV, 
$9\degr\times 9\degr$). Both data sets have been used to produce the source 
light curves in different energy ranges while only FCFOV data have been used for 
spectral extraction because of the better signal to noise ratio.
 
Since the IBIS spectra do not show any substantial variability during 2003 April 
14--21, we have integrated over this observation period. The JEM-X spectra were 
derived with the ISDC OSA 5.0, using the latest available spectral matrices and 
extracting the source spectrum with the fixed position of the source. While IBIS 
provides a large FOV, $>30\degr$, that of JEM-X is narrower ($>10\degr$), thus 
providing only a partial overlap with the high energy detector. In general, the 
effective JEM-X exposure is $\sim$15\% of that of IBIS. Therefore, a 
combined JEM-X/IBIS spectrum was possible only for a small fraction of the data. 

We have extracted the PCA spectrum and the light curve from the {\it RXTE\/} 
public data 
archive\footnote{http://heasarc.gsfc.nasa.gov/docs/xte/xte\_public.html.}. When 
spectral data are obtained in the same time interval from more than one 
instrument, we allowed a free relative overall normalisation constant for each 
instrument with respect to IBIS.

\section{Results}

\subsection{Time evolution}\label{time}

IGR J17091--3624 is a very faint source, which is often below the detection 
limits of the IBIS and JEM-X in a single 2000-s SCW ($\sim$5 mCrab in the 
20--100 keV IBIS energy range, and $\sim$23 mCrab in the 3--20 keV JEM-X energy 
range). Furthermore, most of the JEM-X data were not in its FCFOV, which is 
partly due to the pointing algorithm of the GCDE. As a result, the JEM-X 
effective exposure was $\sim$10 times shorter than that of IBIS.

\begin{figure}
   \centering
 \includegraphics[angle=0, scale=0.8]{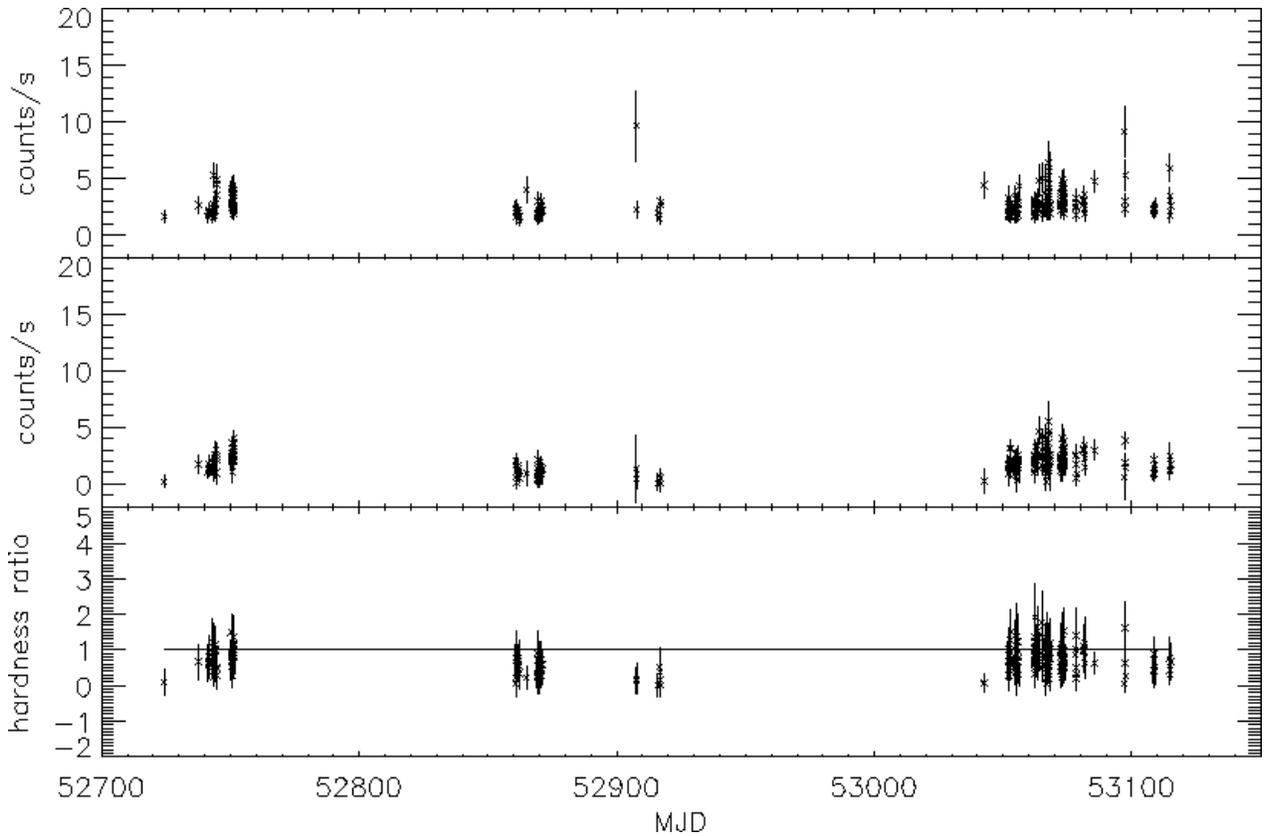}
      \caption{The 20-40 keV and 40-100 keV light curves (the top and middle panel, respectively) and the corresponding hardness ratio (the bottom panel), with the solid line representing a value of to 1.}
         \label{lc_20_50}
	 \label{lc_50_150}
	 \label{HR}
\end{figure}

We have followed the IBIS evolution of the source flux from the first {\it 
INTEGRAL\/} detection of the source. First, a flux increase took place in the 
first SCWs of revs.\ 61--63 (2003 April 14--21), as can be seen in Figure 
\ref{lc_20_50}. We have also found a very faint signal in the data of rev.\ 60. 

We have analysed the data until rev. 185 (2004 April 19--20), as later observations 
do not show any signal in the direction of the source at least until rev.\ 232 
(2004 September), the current limit of the public data set. We have produced 
IBIS light curves in two different energy bands, 20--40 and 40--100 keV (the 
top and middle panel, respectively, of Figure \ref{lc_20_50}). 
The light curves are derived from the flux and variance maps for each individual 
SCW by extracting values directly from the derived source positions. The flux maps 
are already corrected for off-axis effects so this correction is automatically 
included in the light curves. Differences in the size of error the bars on 
individual points are due to both different SCW exposures and different source locations 
within the field of view. The bottom panel of Figure~\ref{HR} shows the hardness ratio, 
$HR\equiv R_{40-100}/R_{20-40}$, where $R$ is the count rate. During MJD 52860--52924 
(2003 August--October), the source was rather faint, especially in the 40--100 
energy range, and the hardness ratio shows a weak indication of softening 
corresponding to 25\%.

The {\it RXTE\/} observation lasted only 2 ks, so that it has not been possible to extract information on the long-term temporal behaviour. A detailed timing analysis of the 2-ks data set is given by Lutovinov \& Revnivtsev (2003). 

For the third observation period, during MJD 53054--53097 (2004 February--April), only IBIS data were available. The light curves are characterised by an increased flux with respect to the previous periods together with a faint tendency of source hardening.

\subsection{Spectral evolution of IGR J17091--3624}
\label{spectral}

We have studied the spectral behaviour separately for 3 epochs:

1) revs.\ 61--63 (2003 April 15--21, $\sim$15 ks), which is the first {\it INTEGRAL\/} observation combined with the {\it RXTE}/PCA data;

2) revs.\ 100--119 (2003 August--October, $\sim$77 ks), during which the IBIS spectrum softened, and which includes the only JEM-X detection.

3) revs.\ 165--179 (2004 April, $\sim$77 ks), containing only the IBIS data.

The data sets have been fitted with XSPEC 11.3.1. Several models have been tested for each of the data sets. In our fits, we have confirmed the constraint found by Lutovinov \& Revnivtsev (2003), $N_{\rm H}<10^{22}$ cm$^{-2}$, which is also in agreement with the Galactic column density along the direction to the source, $\simeq 8\times 10^{21}$ cm$^{-2}$ (http://heasarc.gsfc.nasa.gov). 

\begin{figure}
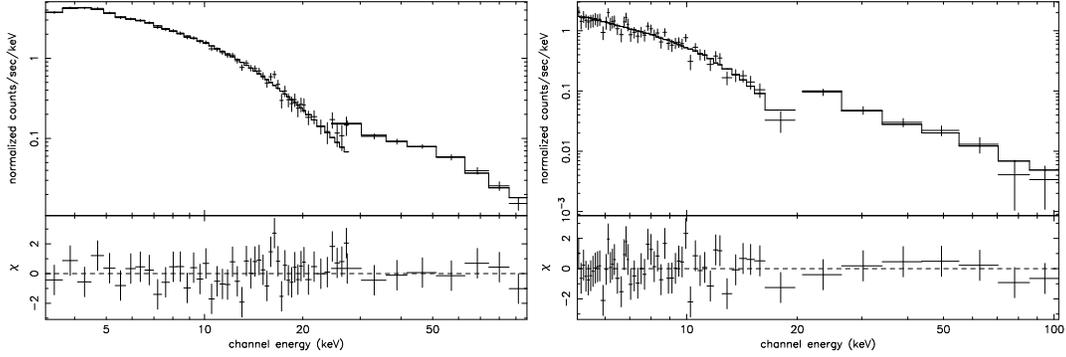

   \centering
   \includegraphics[angle=-90, scale=0.3]{fig3.ps}
 \includegraphics[angle=-90, scale=0.3]{fig4.ps}
      \caption{Left panel: the {\it RXTE}/IBIS count spectra for revs.\ 61--63; Right panel: the JEM-X/IBIS count spectra for revs.\ 100--119. } 
         \label{counts}
  \end{figure}

\begin{figure}
     \centering
  \includegraphics[angle=90, scale=0.7] {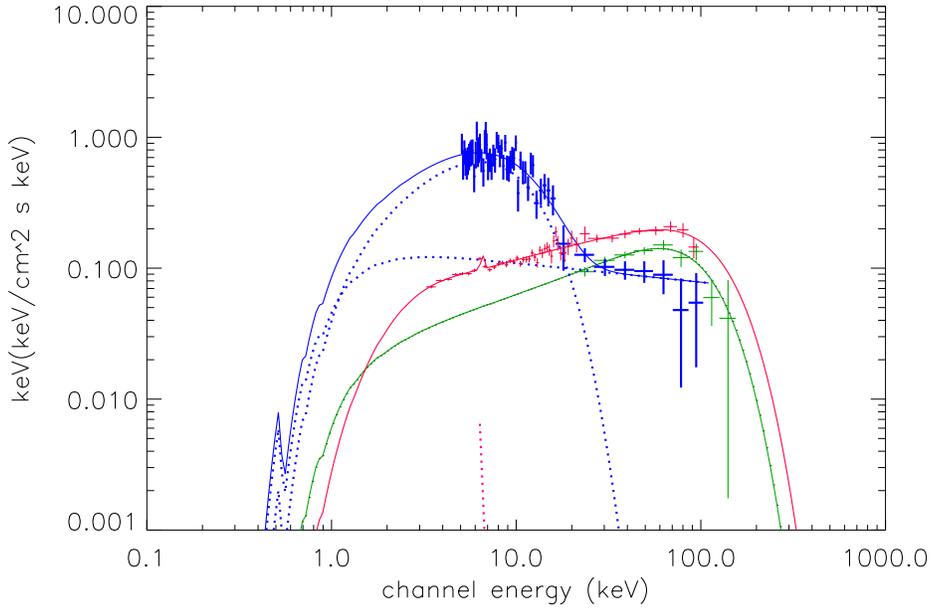}
     \caption{The unfolded spectra showing the source evolution during the {\it INTEGRAL\/} and {\it RXTE\/} observations. The red spectrum (PCA/IBIS) corresponds to revs.\ 61--63, the blue spectrum (JEM-X/IBIS) corresponds to revs.\ 100--119, and the green spectrum (IBIS) corresponds to revs.\ 164--179. The Fe $K_{\alpha}$ line, shown by the red dotted curve, is due to the Galactic ridge emission. The blue dotted curves denote the disk blackbody and power-law components of the models.}
         \label{spectra}
   \end{figure}

For revs.\ 61--63, the relatively high  brightness and the lack of variability have  allowed us to obtain a combined IBIS/PCA spectrum, (see Figure~\ref{spectra}). We use two thermal Comptonization models, {\sc comptt} \cite{Tit} and {\sc compps} \cite{Pou}, both assuming a slab geometry. The {\sc compps} model is an accurate iterative-scattering model, in which subsequent photon scatterings are directly followed. It has been extensively tested against Monte Carlo results, e.g., by Zdziarski, Poutanen \& Johnson (2000). On the other hand, {\sc comptt} is based on an approximate solution of the kinetic equation with some relativistic corrections, and the resulting spectra are also only approximate (see the appendix of Zdziarski, Johnson \& Magdziarz 1996). For the {\sc compps} model, we assumed a viewing angle of $60\degr$ and the source of seed photons to be at the slab bottom (the model geometry parameter was set equal to 1) with a blackbody distribution at $kT_{\rm bb}=0.1$ keV. For the {\sc comptt} model, we used Wien seed photons with the temperature as above. The free parameters of each model are the electron temperature, $kT_{\rm e}$, the Thomson optical depth, $\tau$, and the normalization, $N$. There is no evidence of a Compton reflection component, and the weak fluorescent Fe K$\alpha$ line present in the spectrum appears mostly due to the Galactic ridge emission \cite{RXTE}. Table~\ref{61_63}, the left panel of Figure~\ref{counts} and the red curve in Figure~\ref{spectra} show our results. 

 The results corresponding to revs.\ 100--119 consist of the combined JEM-X (6 SCWs) and ISGRI average spectra. Given the difference in the JEM-X and ISGRI exposures, we have also compared the ISGRI count rate for the 6 SCWs for which the JEM-X spectrum was obtained, and found it is almost the same as the count rate averaged over the entire period of revs.\ 100--119. We have analysed in detail the  JEM-X data in order to exclude any possible contamination from close by sources, being IGR J17091-3624 located in a crowed region near the galactic center. 

  The source shows a bright component in soft X-rays, followed by a weak high-energy tail. Thus, we have used the model consisting of a disk blackbody ({\sc diskbb}, Mitsuda et al.\ 1984) and a power law spectrum. Table~\ref{100_119} summarizes our results, and the spectrum is shown in the right panel of Figure~\ref{counts} and in the blue curve in Figure~\ref{spectra}.

We clearly see a transition from the hard to the soft state from the first to the second epoch. After this, the source returns to the hard state, as shown by the IBIS data for revs.\ 164--179, green spectrum in Figure~\ref{spectra}. The epoch 3 IBIS data are fitted by {\sc comptt} with $kT_{\rm e}= 21^{+4}_{-4}$ keV and $\tau= 2.7^{+1.2}_{-0.5}$; these values are similar to that of epoch 1 (Table~\ref{61_63}). 

The spectral evolution of IGR J17091--3624 shows both similarities and 
differences with respect to that of another black-hole candidate observed by 
{\it INTEGRAL\/}, IGR J17464--3213 \cite{Cap,JOINET}. Therefore, we report here 
the spectra and fit results for that object as well. Table~\ref{1746} shows fit 
results with both {\sc comptt} and {\sc compps} for the hard state and Figure 
\ref{comparison} shows a comparison of its hard and soft states with those of 
IGR J17091--3624.

\section{Discussion and conclusions}

We have followed the evolution of an outburst of  IGR J17091--3624. Within the course of less than one year, the source went from the hard to the soft state and then back to the hard state. 

Although Markoff et al.\ (2005) have recently suggested that the hard state 
emission could be due to synchrotron self-Compton emission from the base of the 
jet, we have considered here the standard hard-state model with the dominant 
radiative process being thermal Comptonization of disk blackbody photons. An 
intriguing feature of our hard-state spectra is the rather low electron 
temperature, $\sim$20 keV, while the range of temperatures typical for 
black-hole binaries is $\sim$50--100 keV (e.g., Zdziarski \& Gierli\'nski 2004). 
This low temperature is obtained using both the {\sc compps} and {\sc comptt} models, 
and for both occurrences of the hard state, in the rise and decline. Such a low 
temperature may be a characteristic of the hard state of neutron-star low-mass 
X-ray binaries, although at present it is rather poorly constrained (e.g., 
Gierli\'nski \& Done 2002). On the other hand, we might have caught the source 
during transitionary phases, with enhanced Compton cooling due to the increasing 
flux of disk blackbody emission and the correspondingly lower electron 
temperature. 

During the soft state in epoch 2, the peak of the $EF(E)$ emission is at several keV, as compared to the $\sim$100 keV peak in the hard state. There is a high-energy tail, but its weakness does not permit us to put constraints on its origin. 

\begin{figure}
     \centering
  \includegraphics[angle=90, scale=0.7] {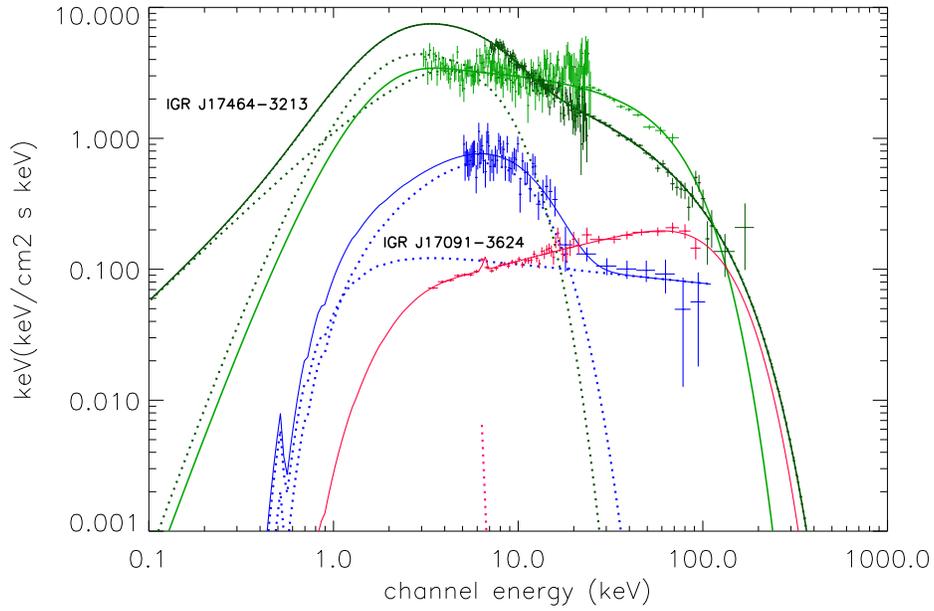}
     \caption{The spectra of IGR J17091--3624 during the rising part of the outburst (red: hard state; blue: soft state), compared with the spectra of IGR J17464--3213 also corresponding to the rising part of its outburst (dark green: soft state; light green: hard state). The blue and red dotted curves denote the same model components as in Fig.\ \ref{spectra}. The dark green dotted curves denote the blackbody and thermal-Compton components of the corresponding fit.}
         \label{comparison}
   \end{figure}

In Section \ref{spectral}, we have also compared the evolution of IGR 
J17091--3624 with that of another transient source observed by {\it INTEGRAL}, 
IGR J17464--3213. This is a black-hole candidate discovered by {\it HEAO 1\/} in 
1977, named then H1743--322. An outburst of that source was observed during the 
GCDE in 2003 \cite{Cap}. Both sources show similar spectral evolution 
during outbursts of similar duration. However, as shown in Figure\ 
\ref{comparison}, the hard-state spectrum of IGR J17464--3213 is substantially 
softer than that of IGR J17091--3624, although both sources share the relatively 
low electron temperature of $\sim$20 keV. Furthermore, the relative 
normalisation of the observed hard and soft state spectra is different. 
Curiously, the high-energy parts ($\ga$30 keV) of both spectra of IGR 
J17464--3213, are virtually identical. The various amplitudes of the hard and 
soft state spectra are often seen in X-ray transients (e.g., Zdziarski et al.\ 
2004), and are due to the hysteretic behaviour of the accreting systems. 

IGR J17091-3624 is located $10\degr$ from  the Galactic Centre  region, where  
the highest concentration of LMXRBs in the Galaxy is present \cite{White}. The 
model unabsorbed bolometric (0.01--500 keV) luminosities in the hard (revs.\ 
61--63) and soft state (revs.\ 100--119), assuming a distance of 8.5 kpc and 
isotropy, are $L\simeq 1\times 10^{37}$ erg s$^{-1}$ and $L\simeq 2\times 
10^{37}$ erg s$^{-1}$, respectively. These values are not far from those 
characteristic to the hard and soft states of black-hole binaries (e.g., 
Zdziarski \& Gierli\'nski 2004), although the neutron-star nature of the source 
cannot be excluded. On the other hand,  the {\it RXTE\/} power spectrum, without 
QPO--like features and with band-limited noise \cite{RXTE}, commonly observed in black-hole binaries in the low/hard spectral state, as well as the probable 
correlation with a radio source \cite{Pand,Atel3} may also argue for the  
black-hole nature of this source. 

\begin{acknowledgements}

We acknowledge the ASI financial/programmatic support via contracts I/R/046/04. 
We thank the unknown referee for his/her useful comments, which helped us to 
improve the quality of this work. We also thank Catia Spalletta for the careful 
editing of the manuscript and Memmo Federici for supervising the {\it 
INTEGRAL\/} data analysis. AAZ has been supported by grants 1P03D01827, 
1P03D01128, 4T12E04727.

\end{acknowledgements}

\clearpage

\clearpage

\begin{table}
\begin{center}
\caption{The log of the {\it INTEGRAL\/} observations of IGR J17091--3624 used for the spectral fits.}
\label{integral_log}
\begin{tabular}{lccc}
\hline
\hline
Revolution & Start Date & End Date & ISGRI exposure [ks] \\
\hline
61--63$^{\mathrm{a}}$   &  2003-04-12 &  2003-04-21 & 15\\
100--119 &  2003-08-10  & 2003-10-04 & 77\\
164--185 &  2004-02-17  & 2004-04-02 & 77\\
\hline
\end{tabular}
\begin{list}{}{} \item[$^{\mathrm{a}}$]{Including the 2-ks {\it RXTE\/} observation on 2003-04-20.}
\end{list}
\end{center}
\end{table}

\begin{table} 
\begin{center}
\caption{The fit parameters to the PCA/IBIS spectrum of revs.\ 61--63 with the {\sc comtt} and {\sc compps} models (see the red spectrum in Figure \ref{spectra}).
\label{61_63}
} 
\begin{tabular}{lccccc}
\hline
\hline
model$^{\mathrm{b}}$ &  $kT_{\rm e}$& $  \tau$&  $\chi^2$/$\nu$&   d.o.f &   $F$(3--100 keV)\\
   &  keV &   &   &   &    erg cm$^{-2}$ s$^{-1}$ \\
\hline 
    {\sc comptt} &  $24^{+9}_{-4}$& $2.1^{+0.3}_{-0.5}$& 0.93& 54& $2.5\times 10^{-10}$\\
    {\sc compps} &  $31^{+12}_{-4}$ &  $2.3^{+0.2}_{-0.4}$ &  0.96& 54 &  $2.4\times 10^{-10}$ \\
\hline
\end{tabular}
\end{center}
\begin{list}{}{} \item[$^{\mathrm{b}}$]{$N_{H}<10^{22}$}
\end{list}
\end{table}

\begin{table}
\begin{center}
\caption{The fit parameters to the JEM-X/IBIS spectrum of revs.\ 100--119 with the disk blackbody and a power law model (see the blue spectrum in Figure \ref{spectra}).}
\label{100_119}
\begin{tabular}{lccccccc}
\hline
\hline
 Model & $kT_{\rm in}$ & $N_{\rm disk}$ & $\Gamma$ & $N_{\rm pl}$$^{\mathrm{c}}$ & $\chi^2$/$\nu$ & d.o.f. & $F$(3--100 keV)\\
&keV& & & & & & erg cm$^{-2}$ s$^{-1}$ \\
\hline
{\sc diskbb+pow} & $2.6^{+0.2}_{-0.2}$ & $1.9^{+0.7}_{-0.6}$ & $2.1^{+0.3}_{-0.4}$ & $0.2^{+0.4}_{-0.1}$ & 1.0 & 53 & $2 \times 10^{-9}$\\
\hline
\end{tabular}
\end{center}
\begin{list}{}{} \item[$^{\mathrm{c}}$]{${\rm keV}^{-1}\,{\rm cm}^{-2}\,{\rm s}^{-1}$ at 1 keV.}
\end{list}
\end{table}

\begin{table}
\begin{center}
\caption{The fit parameters to the  hard-state spectrum of the black-hole candidate IGR J17464--3213 (= H1743--322, Capitanio et al.\ 2005) with the {\sc comtt} and {\sc compps} models (see the light green spectrum in Figure \ref{comparison}).} \label{1746}
\begin{tabular}{lccccc}
\hline
\hline
model &  $kT_{\rm e}$& $  \tau$& $\chi^2$/$\nu$&   d.o.f &   $F$(3--100 keV) \\
   &  keV &   &    &   &    erg cm$^{-2}$ s$^{-1}$ \\
\hline 
{\sc comptt} &  $18^{+4}_{-2}$ & $1.3^{+0.2}_{-0.4}$ &1.05&119 & $6\times 10^{-9}$\\
    {\sc compps} &  $20^{+3}_{-1}$ & $2.0^{+0.1}_{-0.2}$ & 1.04& 119 & $4 \times 10^{-9}$\\
\hline
\end{tabular}
\end{center}
\end{table}

\end{document}